\documentclass[aps,a4paper,superscriptaddress,nofootinbib,tightenlines
,floats,amsfonts,amssymb]{revtex4}

\usepackage{graphicx}
\usepackage{latexsym}

\input epsf

\def\H{\mathcal{H}}
\def\L{\mathcal{L}}
\def\rd{\mathrm{d}}
\def\d{\partial}

\def\cos{{\rm cos}}

\def\be{\begin{equation}}
\def\ee{\end{equation}}
\def\bea{\begin{eqnarray}} 
\def\eea{\end{eqnarray}} 
\def\half{\frac{1}{2}}
\def\quarter{\frac{1}{4}}
\def\simleq{\; \raise0.3ex\hbox{$<$\kern-0.75em
      \raise-1.1ex\hbox{$\sim$}}\; }
\def\simbeq{\; \raise0.3ex\hbox{$>$\kern-0.75em
      \raise-1.1ex\hbox{$\sim$}}\; }

\begin{document}

\title{Nielsen-Olesen strings in Supersymmetric models}
\author{M. Pickles}
\affiliation{DAMTP, Centre for Mathematical Sciences, Cambridge
University, U.K.}

\author{J. Urrestilla}
\affiliation{Department of Theoretical Physics, UPV-EHU, Bilbao, Spain}

\date{\today}

\begin{abstract}
We investigate the behaviour of a model with two oppositely charged
scalar fields. In the Bogomol'nyi limit this may be seen as the scalar
sector of $N\!\!=\!\!1$ supersymmetric QED, and it has been shown that cosmic
strings form. We examine numerically the model out of the Bogomol'nyi
limit, and show that this remains the case. We then add supersymmetry-breaking
mass terms to the supersymmetric model, and show that strings still survive.
 Finally we consider the
extension to $N\!\!=\!\!2$ supersymmetry with supersymmetry-breaking mass terms, and show that this leads to the formation of stable cosmic
strings, unlike in the unbroken case.

\end{abstract}

\maketitle

\section{Introduction}

Supersymmetric (SUSY) theories often have a continuous degeneracy of
inequivalent ground states. Classically this corresponds to flat
directions of the scalar potential, which may arise from abelian
theories when the gauge symmetry is broken by a Fayet-Iliopoulos
D-term. Recently, these have undergone intense study, mostly 
in the context of effective actions for SUSY non-abelian gauge theories 
and confinement \cite{dterm}.

We naturally expect the formation of cosmic strings when the D-term breaks 
the gauge symmetry
 \cite{PRTT96, DDT97}, but the
presence of the flat directions means that the situation is more
complicated. It is not immediately obvious which vacuum states among
the degenerate vacua will give rise to stable string
solutions. 

The situation has been investigated  in \cite{PRTT96} where a
model with two oppositely charged chiral multiplets and zero
superpotential is discussed. They show that one particular vacuum is
favoured, and that a
Nielsen-Olesen (NO) \cite{NO73}  type cosmic string can form in one of the scalar
fields, the other field being zero everywhere. It was later
shown in \cite{ADPU01} that the crucial role in this vacuum selection
effect is that played by the gauge field. The vaccum is effectively
 chosen so that 
the vector mass is minimized,
 leading to a lower energy for the core of the vortex.
The authors in \cite{ADPU01} also proved that the preferred cosmic string was stable,
using a Bogomol'nyi type of argument.

In this paper we consider  the scalar sector of
this theory outside the Bogomol'nyi bound to study whether the result mentioned above
holds as well for other values of the parameters.
 Although it no longer
corresponds to the simplest SUSY model, it still possesses
flat directions, and may be viewed  as
illustrating the behaviour of more general SUSY theories
with flat directions. An analytical approach is not possible out of the Bogomol'nyi bound, so we carry out
 numerical simulations of the fields
to see if there is still any evidence of vacuum selection
occurring. In fact the simulations suggest that, for all the range of parameters considered, the system
chooses one out of all the possible minima of the flat directions and forms NO strings.

We then study the fate of the different defects formed in the $N\!\!=\!\!1$ and $N\!\!=\!\!2$ supersymmetric 
QED models considered in \cite{ADPU01} in which we break supersymmetry by the addition of soft
supersymmetry-breaking mass terms to the scalars. Then, the degeneracy in the minima of the potential is lifted, and we 
have no flat directions. For the $N\!\!=\!\!1$ case, the NO string which is present without mass terms can survive SUSY breaking 
for some values of the masses. 

The situation is different for the $N\!\!=\!\!2$ model. The system was first considered in \cite{GMV96}, in which it was claimed  
that the model accommodated strings that could confine magnetic charges.
 However, \cite{ARH98} showed that the strings
were not stable, and so could not confine. A further step attemping to get stable strings by adding a Fayet-Iliopoulos term \cite{ADPU01} proved to be insufficient.
Although there is 
a vacuum selection effect in the model, it is not enough to form {\it stable} cosmic
strings. The strings formed are semilocal \cite{VA91}, and are only
neutrally stable, in the sense that small perturbations tend to cause the 
string core to expand indefinitely \cite{leese}. We will show that with the 
additional supersymmetry breaking mass terms, stable cosmic strings are 
possible.


\section{A scalar model out of the Bogomol'nyi bound}

The model we consider has two complex scalar fields $\phi_+$ and $\phi_-$ of
opposite charges, coupled to a $U(1)$ vector field. The Lagrangian is
of the form

\bea
\L=|D_\mu\phi_+|^2+|D_\mu\phi_-|^2-\quarter
F_{\mu\nu}F^{\mu\nu}-\frac{\lambda}{2}
\left(|\phi_+|^2-|\phi_-|^2-\eta^2\right)^2\,, \label{lag}
\eea

where $A_{\mu}$ is the $U(1)$ gauge field and, as usual,
\bea
D_\mu\phi_+&=&\left(\d_\mu+iqA_\mu\right)\phi_+\,;\nonumber\\
D_\mu\phi_-&=&\left(\d_\mu-iqA_\mu\right)\phi_-\,;\nonumber\\
F_{\mu\nu} &=&\d_\mu A_\nu-\d_\nu A_\mu\,.
\eea

The above can be regarded, in the Bogomol'nyi limit $\lambda=q^2$, as the
bosonic part of a four-dimensional
 model with two $N\!\!=\!\!1$ chiral superfields ($\Phi_+$,
$\Phi_-$) of opposite charges $\pm q$ coupled to a $U(1)$ vector
multiplet, with zero superpotential and a Fayet-Iliopoulos term
proportional to $\eta^2$.

The vacuum manifold of this model is given by
\be
|\phi_+|^2-|\phi_-|^2=\eta^2\,,
\ee
with solutions
\be
|\phi_+|=\eta \cosh u\equiv v_+\,, \qquad |\phi_-|=\eta \sinh u\equiv v_- \,, 
\label{vacuum}
\ee
where $u$ parametrises the moduli space. When $u=0$, the field
$\phi_-$ is zero, and formation of the usual NO 
string \cite{NO73} in the field $\phi_+$ may occur, since we are back at the
 abelian Higgs model.

By considering the energy per unit length of straight static vortices,
we find that the conditions for finite energy impose some relations between
 the fields:  since we are looking to construct
 cylindrically symmetric configurations,
fields with charge $\pm q$ wind as $e^{\pm
iqn\theta}$, and the gauge field tends to a constant $A_\theta\to-n$.
 Bearing this in mind,
we use the ansatz
\be
\phi_+=f_+(r)e^{iqn\theta}\,,\qquad\phi_-=f_-(r)e^{-iqn\theta}e^{i\Delta}\,,\qquad A_\theta\rd \theta=a(r)\rd \theta\,,
\ee
where $f_\pm$ and $a$ are real functions, and $\Delta$ is a real
constant. $f_\pm(0)=a(0)=0$ for the functions to be well-defined, and $a(r\to\infty)=-n$ by the previous energy arguments.
Such considerations also tell us that the scalar fields should
go to a minimum of the potential energy as $r\to\infty$, and hence the
boundary conditions for the $f_\pm$ functions are $f_\pm(r\to\infty)=v_\pm$.
It is possible to construct vortices tending to any of the
vacua (\ref{vacuum}) as $r\to\infty$.

However, it has been shown \cite{PRTT96,ADPU01} that when the Bogomol'nyi
limit of this system ($\beta=\lambda/q^2=1$) is considered,
the only static solutions are those with $u=0$. 
After symmetry breaking, apart from the massless fields, 
we are left with a massive vector field ($m_v^2=2q^2\eta^2\cosh\,2u$) and a massive scalar field 
($m_s^2=2\lambda\eta^2\cosh \,2u$). The system chooses $u=0$ in order to
minimize the vector mass.

Thus NO strings
may form only in the $\phi_+$ field, since $f_-(r)=0$. 
In the following section we would like
to move out of the Bogomol'nyi limit to see whether 
$u=0$ is still preferred by the system. 
In fact, as explained in \cite{PRTT96}, the dynamics of the magnetic
core and those of the scalar fields outside the core are effectively
decoupled. Far from the core, the magnetic field is very small, and the scalar
 fields lie on the vacuum manifold.
Hence, scalar fields
are free to move along the moduli space with no
appreciable cost in energy, so one might presume that some $\phi_-$
lumps would be formed far from the core of the string and this  might
have some cosmological consequences.


\section{Numerical simulations}

We use numerical simulations of a discretized version of the model
described by (\ref{lag}) in order to study its behaviour out of the Bogomol'nyi
bound. We firstly set 
$\eta=q=1$ by rescaling the fields and spacetime coordinates, and then
we use techniques borrowed from Hamiltonian lattice gauge theory
(see for instance \cite{MMR88}) to perform the simulations. 

The lattice link and plaquette operators are defined as
\bea
& &U_i(x)=e^{-ilA_i(x)}\,;\\
& &Q_{ij}=U_j(x)U_i(x+x_j)U^\dagger_j(x+x_i)U^\dagger_i(x)\,,
\eea
respectively, where $l$ is the lattice spacing, the label $i$ takes the values
$1,2,3$ corresponding to the three spatial dimensions, and $A_i$ are the
gauge fields. By $x+x_i$, we denote the nearest lattice point in the $i$ direction from $x$. 
 The plaquette operators are related to the gauge field strength.
The lattice link operator, on the other hand, is used to define
discrete covariant derivatives
\bea
D_i\phi_+(x)&=&\frac{1}{l}\left(U_i^\dagger(x)\phi_+(x+x_i)-\phi_+(x)\right)
\,;\nonumber\\
D_i\phi_-(x)&=&\frac{1}{l}\left(U_i(x)\phi_-(x+x_i)-\phi_-(x)\right)\,.
\eea
We can then obtain a discretized Hamiltonian
density corresponding to the Lagrangian (\ref{lag})
\bea
\H&=& |\Pi_+|^2+|\Pi_-|^2+\half
E_iE^i+\frac{\beta}{2}\left(|\phi_+|^2-|\phi_-|^2-1\right)^2 \nonumber\\
& &+|D_i\phi_+|^2+|D_i\phi_-|^2+\frac{1}{2l^4}\sum_{i\ne
j}\left(1-{\rm Re}(Q_{ij})\right) \,,
\label{h}
\eea
where $\Pi_\pm$  and $E^i$ are the conjugate momenta of $\phi_\pm$ and
$A_i$  respectively, and $\beta\!\!=\!\!\lambda/q^2$ 
($\beta\!\!=\!\!\lambda$ in rescaled units). In the
$l\to0$ limit, the continuum Hamiltonian for 
the system is recovered.

One interesting property of (\ref{h}) is that it
is gauge invariant, in the sense that a general $U(1)$ transformation
$\Lambda(x)$ for which the fields transform as
\bea
\phi_+(x)&\to&\Lambda(x)^\dagger\phi_+(x)\,;\nonumber\\
\phi_-(x)&\to&\Lambda(x)\phi_-(x)\,;\nonumber\\
U_i(x)&\to&\Lambda(x)U_i(x)\Lambda^\dagger(x+x_i)\,,
\eea
leaves the Hamiltonian invariant.

In order to get the Hamiltonian in the form (\ref{h}) a gauge choice
has been made, namely, $A_0=0$. The equation of motion corresponding
to $A_0$ is equivalent to Gauss's law
\be
il\left(\phi_+^\dagger(x)\Pi_+-\Pi_+^\dagger\phi_+(x)+\Pi_-^\dagger(x)\phi_-(x)-\phi_-^\dagger(x)\Pi_-(x)\right)=\sum_i\left(E^i(x)-E^i(x-x_i)\right)\,.
\label{gauss}
\ee
The system is  geometrically forced  to preserve Gauss's
law, and,  if the initial conditions satisfy relation (\ref{gauss}), subsequent evolution by the equations of motion coming from (\ref{h}) will also satisfy (\ref{gauss}).

The gauge invariant dissipative terms $\gamma\Pi_\pm$
and $\gamma E^i$ were added to the Hamiltonian equations of motion for $\phi_\pm$ and $A_i$
respectively.  These obey
Gauss's law, and thus the system as a whole will still obey Gauss's
law as well. $\gamma$ is a free parameter in the model, but simulations using different values of
$\gamma$ show qualitatively similar behaviour.

The
equations of motion coming from the Hamiltonian (\ref{h}) were simulated numerically on
a $64^3$ lattice.
We are interested in the role played by the $\phi_-$ field in the dynamics
of the system, not in the details of the phase transition itself.
Furthermore, previous
work concerning the numerical simulation of defects showed that 
 the formation of defects depends 
on the later interplay
between scalar and gauge fields, and is  not actually very sensitive to the initial
configuration used \cite{ABL96,UABL01}.
We did, however, use different initial conditions to verify
 that the result was not
dependent on how the system was set. In fact,
all simulations using different initial conditions give rise to
qualitatively the same results.
The first time steps are a transient period during which the system
loses energy rapidly, and relaxes.  Then, the interaction between
gauge and scalar fields leads to the formation of defects.

All the results shown in this paper are obtained by using 
 $\phi_\pm=0$, $A_\mu=0$ and $\dot{A}_\mu=0$  and  random initial velocities
for the scalar fields as the initial configuration. 
This choice of conditions obeys Gauss's law, and hence
we have guaranteed that Gauss's law holds throughout the
simulation. 
Computation of
Gauss's law during the evolution of different initial conditions was
used to ensure the stability of the code.

\section{Numerical results}

\begin{figure}[!htb]
\centering
\leavevmode
\epsfysize=5cm \epsfbox{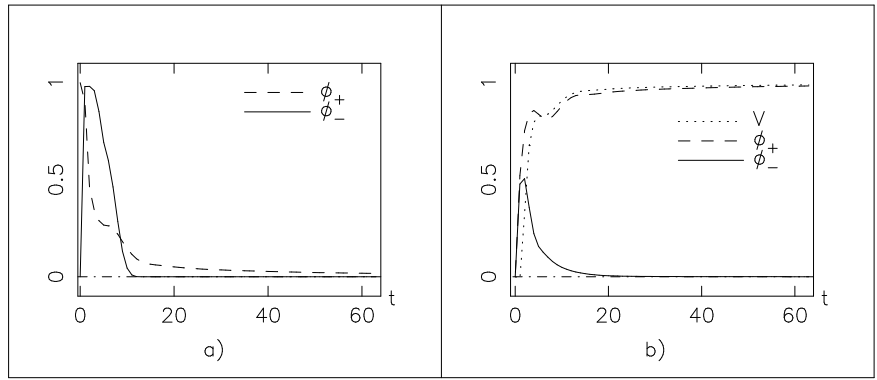}\\
\caption[ene]{\label{ene} Representation of the results of a numerical
simulation performed in a $64^3$ lattice, with $\beta\!\!=\!\!0.3$ and
$\gamma\!\!=\!\!0.5$. Figure a) shows the number of lattice sites with
$|\phi_+|\!\!<\!\!0.75$ (dashed line) and $|\phi_-|\!\!>\!\!0.1$ (continuous line)  as
a fraction of the total number of sites. Figure b) shows the sum of
the modulus of each of the fields at every site (as a fraction of the
total lattice points): $\sum_x |\phi_+|$ (dashed line), $\sum_x |\phi_-|$ (continuous
line). There is also a representation of the quantity
$1-\sum_x V(x)/ V_{\rm max}$, where $V(x)$ is the potential energy
(dotted line).}
\end{figure}

Figure~\ref{ene} is a schematic example of the outcome of the
simulations. It can be  seen  that there exists a transitional period
 during the
early timesteps, when the system is trying to get a physical
configuration by dissipating energy. Both scalar fields acquire
non-zero values at this time, and in particular $\phi_-\!\!\ne\!\!0$.
Some of the initial kinetic energy is transformed 
to potential energy, but there is no hint of structure of any kind yet.
At the end of the transitional period, $|\phi_-|$ decreases rapidly until
$|\phi_-|\!\!\sim\!\!0$.

In a) we have plotted the number of sites in the lattice with
$|\phi_+|\!\!<\!\!0.75$, and also the number with $|\phi_-|\!\!>\!\!0.1$, as a
fraction of the total number of lattice sites. 
After the transitional period ($t\!\!\sim\!\! 15$), there are no points on the
lattice with $|\phi_-|\!\!>\!\!0.1$,
 whereas the field $\phi_+$ has modulus greater
than $0.75$ almost everywhere. The scalars fields should tend to
minimize the potential energy, and  according to these results, it seems that
only $\phi_+$ is involved in achieving this. This is strongly
suggestive of the idea that, out of all the possibilities, the system
has chosen $u=0$ as the preferred point in the flat directions (\ref{vacuum}).

In figure~\ref{ene} b), we have plotted the quantity $1-\sum_x\frac
{V(x)}{V_{\rm max}}$, where $V(x)$ is the potential energy at lattice
point $x$, and $V_{\rm max}$ is defined as the total potential energy
calculated in the case when $\phi_\pm=0$ everywhere. 
$\sum_x |\phi_+(x)|$ and $\sum_x|\phi_-(x)|$ are also plotted 
as fractions of the total number of lattice points.
In this figure we see that, as in a), after the transitional period
$\phi_-$ tends quickly towards zero, while $|\phi_+|\sim 1$ almost
everywhere. The potential energy mimics the behaviour of $\phi_+$, and
ignores the presence of $\phi_-$, suggesting once again that the actual
minimization of the potential energy is due to the field $\phi_+$
alone. 

\begin{figure}[!htb]
\centering
\leavevmode
\epsfysize=5cm \epsfbox{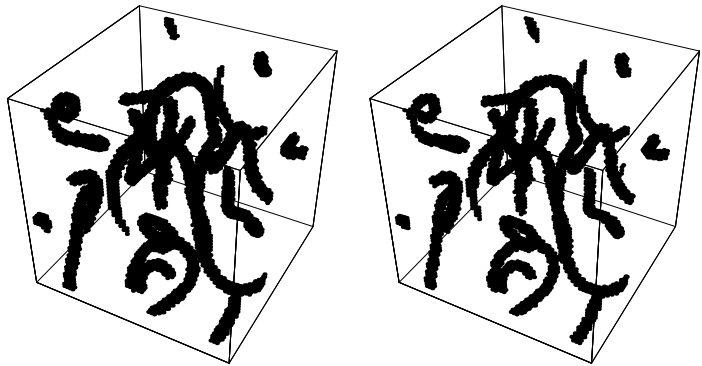}\\
\caption[flux]{\label{flux} Isosurface representation of the modulus
of $\phi_+$ (left), where $|\phi_+|\sim 0.75$ is shown, and
magnetic field (right), where $\sqrt{\half F_{ij}F^{ij}}\sim 0.25$ of the
measured maximum is displayed, for a $64^3$ simulation with $\beta=0.3$, $\gamma=0.5$ at $t=40$.}
\end{figure}

The system is not exactly at the minimum of the potential everywhere,
as there are a number of points in the lattice at which
$|\phi_+|\!\!<\!\!0.75$. This indicates that there are points at the false vacuum ($\phi_+\!\!=\!\!\phi_-\!\!=\!\!0$), and there 
may be strings forming in
the system. In figure~\ref{flux} we give a representation of the
magnetic field, $\sqrt{\half F_{ij} F^{ij}}$, and the modulus of the field $\phi_+$
 at time $t\!\!=\!\!40$, after
the transient period is over. It is clear that the points where
$|\phi_+|\!\!<\!\!0.75$ form structures, namely strings, and that the areas in which 
the  magnetic field is concentrated corresponds exactly to these points. A 
similar plot using $|\phi_-|$
would show that $|\phi_-|\!\!\sim\!\! 0$ everywhere.
The excellent match described above between the magnetic field and $|\phi_+|$ is another
good check of the stability of our code.
We performed several simulations varying $\beta$, from $\beta\!\!=\!\!0.1$ to
$\beta\!\!=\!\!2.0$, which showed the same behaviour.

These results show that after some initial (non-physical) 
period of relaxation, only one of the scalars $(\phi_+)$ becomes dynamical, and
that the other $(\phi_-)$ goes rapidly to zero everywhere, decoupling
from the other fields.
Thus the analytical result obtained for $\beta\!\!=\!\!1$ \cite{ADPU01} has been shown to be
valid for cases with $\beta\!\!\ne\!\!1$. In all cases the system chooses the
vacuum with $u\!\!=\!\!0$, and the field $\phi_-$ is effectively decoupled from the
dynamics.
After this decoupling, 
 we are left with what essentially is an abelian Higgs model 
and therefore the system accommodates stable Nielsen-Olesen
strings, formed by
interaction of the gauge field with $\phi_+$.  

\section{Adding mass terms}

In the previous section we demonstrated that although the vacuum of
the system investigated has flat directions, the only vacuum
configuration appearing in the evolution of a string network is the
one with $u\!\!=\!\!0$, which is a stable cosmic string. We will now proceed to show another situation in which stable
NO strings form in this context.
We first add mass terms to our
Lagrangian (\ref{lag}), obtaining NO vortices again, and then move on
to a more complicated system with $N\!\!=\!\!2$ SUSY, which we will
again break with mass terms. In this second case, when there are no mass
terms, the strings formed are semilocal strings \cite{VA91}, which
in the limit considered are neutrally stable. 
There is a family of such strings,
 all degenerate in energy, with different
widths ranging from that of the NO vortex to an infinite core \cite{H91}.
There exists a zero mode corresponding to changes in the width of the core, 
which when  excited causes any given string in this family to grow 
indefinitely.

Since SUSY is clearly not
manifest at low energies, it is  necessary that it be broken at some scale. Adding these soft
supersymmetry-breaking mass terms to the Lagrangian is a convenient
way of doing so while avoiding the details of
how the breaking actually occurred.

The new Lagrangian, modified from (\ref{lag}), is
\be
\tilde\L=|D_\mu\phi_+|^2+|D_\mu\phi_-|^2-\quarter
F_{\mu\nu}F^{\mu\nu}-\frac{\lambda}{2}
\left(|\phi_+|^2-|\phi_-|^2-\eta^2\right)^2-m_+^2|\phi_+|^2-m_-^2|\phi_-|^2
\,,
\ee 
and the two extrema of the potential are now at 
\be
 \phi_+=  0\, , \qquad  \phi_-=0\,,
\ee
and
\be 
 |\phi_+|^2  =  \eta^2-\frac{m_+^2}{\lambda}\,, \qquad \phi_-=0\,.
\ee

When $\lambda\eta^2\!\!>\!\!m_+^2$, the second case is clearly the
minimum. The degeneracy of the vacuum is lifted, $u=0$ is
selected out of the possible vacua (\ref{vacuum}), and 
we are left with a genuine NO vortex. If the condition 
$\lambda\eta^2\!\!>\!\!m_+^2$ 
does not hold,
the only solution has both fields equal to zero, and no strings form.
Therefore, depending in the value of $m_+$, the system either accomodates a
NO string solution in $\phi_+$, or no string solution is possible at all.
 As in 
the previous case, $\phi_-$ does not play any role in the defect formation.

A more interesting result is obtained when this $N\!\!=\!\!1$ model is
 upgraded 
to $N\!\!=\!\!2$. We consider a model consisting of two $N\!\!=\!\!2$
hypermultiplets $h_a$, where $a\!\!=\!\!1,2$, of opposite charges 
$q_a\!\!=\!\!\pm q$, coupled
to an $N\!\!=\!\!2$ abelian vector multiplet. Such a model was considered in
\cite{GMV96} as a candidate for obtaining magnetic flux confinement in
the low energy limit of type II superstrings compactified on
Calabi-Yau manifolds. The vortex solution obtained in \cite{GMV96} was
found to be unstable \cite{ADH98}, since the $U(1)$ gauge symmetry was
not broken. A new attempt to obtain stable vortices in this model
\cite{ADPU01} added a Fayet-Iliopoulos D-term of the form $\vec{k}\cdot
\vec{D}$ to the system. This gives rise to a term that spontaneously
breaks the gauge symmetry. The defects
obtained in this model were the so-called semilocal strings
\cite{VA91}, a one-parameter set of magnetic "vortices", which
includes the NO vortex. In the limit studied, semilocal strings are
known to spread out to greater and greater radius, as mentioned before, until the string
relaxes to the vacuum \cite{H91}, although the magnetic field remains quantized.

In the present work, we add soft supersymmetry-breaking mass terms to the
Higgs scalars of the model studied in \cite{ADPU01}, and investigate the
resulting vortices. 

Without loss of generality we take
$\vec{k}=(0,0,\frac{\eta^2}{2})$. The energy of the system is
\begin{eqnarray}
E & = & \int\!\! \rd^2 \!x \,\half\left[|D_\mu h_{11}|^2 +
     |D_\mu h_{12}|^2 + |D_\mu h_{21}|^2  +|D_\mu h_{22}|^2 +
     B^2 \right. \nonumber \\
     & & 
       +(H^{\;1}_1-H^{\;2}_2 + \frac{\eta^2}{2})^2
         + (H^{\;1}_2 + H^{\;2}_1)^2+ (i\,H^{\;1}_2 - i\,H^{\;2}_1)^2
     \nonumber\\
& & + m_{11}^2|h_{11}|^2 +m_{12}^2|h_{12}|^2 +m_{21}^2|h_{21}|^2
     +m_{22}^2|h_{22}|^2 \left. \begin{array}{c} \\ \\ \end{array}
     \!\!\!\! \right]\,,
\label{model}
\end{eqnarray}
where $D_{\mu} = \partial_\mu + {i (q_a)} A_\mu$ and 
$H^{\;\;i}_j = - ({q_a}/{2})\,h_{ai}^*h_{aj} $.

The original model had, in addition to the $U(1)$ gauge symmetry, an
effective global $SU(2)$ symmetry of the scalar potential, mixing the pairs of fields
($h_{11}$ and $h_{22}^*$) and ($h_{12}$ and $h_{21}^*$) amongst
themselves. This symmetry was responsible for the appearance of
semilocal strings. It would therefore be expected that the addition of
different mass terms $m_{11} \ne m_{22}$ would lead to the formation
of NO vortices, as the $SU(2)$ symmetry is broken by those terms. 

Parametrising the scalars as $h_{ai}=r_{ai}e^{i\chi_{ai}}$, the
potential takes the form
\bea
V&=&\frac{1}{8}\left(\;\left(-r_{11}^2+r_{21}^2+r_{12}^2-r_{22}^2+\eta^2\right)^2
+4r_{11}^2r_{12}^2+4r_{21}^2r_{22}^2-\right.\nonumber\\
& &\left.8\left(  
 r_{11}r_{12}r_{21}r_{22} \cos(\chi_{11} +
\chi_{22} - \chi_{12} - \chi_{21})\;
\right) \; \right) \nonumber\\
& & +\half\left(m_{11}^2 r_{11}^2+m_{12}^2r_{12}^2+m_{21}^2 r_{21}^2+ m_{22}^2
r_{22}^2\right)\,. 
\eea

In order to obtain the minimum potential we require $\cos(\chi_{11} +
\chi_{22} - \chi_{12} - \chi_{21})$ is maximised, so that  $\chi_{11} +
\chi_{22} - \chi_{12} - \chi_{21}=0$ modulo  $2 \pi$.
The vacuum manifold can then be calculated by varying $V$ over all the
fields, $\frac{\delta V}{\delta r_{ai}}=0$, and we find the following minima
\bea
r_{11}^2&=&\eta^2-2m_{11}^2\, ,\qquad r_{12}=r_{21}=r_{22}=0\,;\nonumber\\
r_{12}^2&=&-\eta^2-2m_{12}^2\, ,\qquad r_{11}=r_{21}=r_{22}=0\,;\nonumber\\
r_{21}^2&=&-\eta^2-2m_{21}^2\, ,\qquad r_{11}=r_{12}=r_{22}=0\,;\nonumber\\
r_{22}^2&=&\eta^2-2m_{22}^2\, ,\qquad r_{11}=r_{12}=r_{21}=0\,,
\eea
together with the maximum
\be
r_{11}=r_{12}=r_{21}=r_{22}=0\,,
\ee
supposing that all masses are different.  We see that solutions with
$r_{12}\ne0$ and  $r_{21}\ne0$ are never valid, so we are left with only
two possible minima. These two minima correspond to NO vortices in
$r_{11}$ or $r_{22}$, and depending on the values of $m_{11}$ and
$m_{12}$ one or the other will form. 

For $m_{11}<m_{22}$ and $2m_{11}^2<\eta^2$ the NO strings will form in the 
$h_{11}$ field; on the other hand, if $m_{22}<m_{11}$ and
$2m_{22}^2<\eta^2$ the NO will be in the $h_{22}$ field. If the masses
are bigger than $\eta^2$, then no vortex solution is
possible. 

A special case is when $m_{11}=m_{22}<\eta^2/2$, since in this
instance the $SU(2)$ symmetry between $h_{11}$ and $h_{22}^*$ is not broken, and we
recover semilocal strings as in \cite{ADPU01}. The minima in this case are
\be
r_{11}^2+r_{22}^2=\eta^2-2 m_{11}^2\,,
\ee
and we have a whole family of solutions \cite{H91}, the NO string among them. 

We then conclude that, depending on the values of the added masses $m_{11}$ and $m_{22}$, we
might  have the case where no strings at all form, the case where we recover semilocal strings, 
or the system might be able to form strings in $h_{11}$ or $h_{22}$
 that are stable NO vortices.


\section{Discussion}

We have seen that, for a general class of scalar fields with flat
directions in the potential, there does seem to be a vacuum selection
effect taking place even when the system is out of the Bogomol'nyi limit.
Furthermore, the vacuum chosen remains the
same as in the Bogomol'nyi limit, regardless of the
value of the parameter $\beta$.
After spontaneous symmetry breaking, defects may form. Due to
this vacuum selection effect only one of the fields becomes
dynamical and this field gives rise to Nielsen-Olesen strings,
whereas the other field decouples.
This is what is seen in simulations. 

On the other hand, for supersymmetric models with flat directions, the
 addition of supersymmetry-breaking mass terms, that arise in the effective
 SUSY action at low energies, lifts these flat directions. In the case that
  $m_{11}\!\!=\!\!m_{22}$, the string is a semilocal string in the
 Bogomol'nyi limit. This solution has the same form as that of the model
 without masses, where there are no stable string solutions. Nevertheless, if
 the masses are different, only one scalar field has a non-zero vacuum
 expectation value. This field then gives rise to a NO cosmic string.

In addition, in supersymmetric theories there will be fermion zero modes in the string cores due to the symmetry between bosons and fermions \cite{DDT97}. 
In the case where the supersymmetry is broken, it may still be
possible to have fermionic zero modes on these strings. The zero modes  give rise to currents that can stabilise string loops. We will investigate this
possibility in a subsequent paper \cite{ADPU02}\,.

\section*{Acknowledgements}
We would like to thank Anne-Christine~Davis and Ana~Ach\'ucarro for helpful
discussions. We thank the University of Leiden, where part of this
work was done, for hospitality. This work is partially supported by 
the ESF COSLAB programme. JU is supported by AEN99-0315, FPA 2002-02037
and  9/UPV00172.310-14497/2002 grants.

\end{document}